\documentclass[fleqn,usenatbib]{mnras}

\usepackage{newtxtext,newtxmath}

\usepackage[T1]{fontenc}

\DeclareRobustCommand{\VAN}[3]{#2}
\let\VANthebibliography\thebibliography
\def\thebibliography{\DeclareRobustCommand{\VAN}[3]{##3}\VANthebibliography}

\usepackage{graphicx}	
\usepackage{amsmath}	

\title[RRAT Astronomy report 2021]{Search for Dispersed Pulses at Declinations from $+56^{\circ}$ to $+87^{\circ}$}

\author[S. A. Tyul'bashev et al.]{
Sergei A. Tyul'bashev,$^{1}$\thanks{E-mail: serg@prao.ru (SAT)}
Marina A. Kitaeva,$^{1}$
Sergei V. Logvinenko,$^{1}$
and Gayane E. Tyul'basheva$^{2}$
\\
$^{1}$ P.N. Lebedev Physical Institute of the Russian Academy of Sciences, Astro Space Center, Pushchino Radio Astronomy Observatory,\\
Radiotelescopnaya 1a, Moscow reg., Pushchino, 142290, Russia \\
$^{2}$ Institute of Mathematical Problems of Biology, brunch of Keldysh Institute of Applied Mathematics,\\  
Vitkevich 1, Moscow reg., Pushchino, 142290, Russia\\
}

\date{December 2021}

\pubyear{December 2021}

\begin{document}
\label{firstpage}
\pagerange{\pageref{firstpage}--\pageref{lastpage}}
\maketitle

\begin{abstract}
The survey of northern hemisphere were made at a frequency of 111 MHz. The total accumulation time at each point of the sky was at least one hour. 75 sources of pulse emission were detected. More then $80\%$ of these sources are known pulsars observed in the side lobes of the antenna. From one to several hundreds pulses were detected in twelve known pulsars. For four pulsars (J0157+6212, J1910+5655, J2337+6151, J2354+6155) the narrowness of the strongest pulses and the ratio of peak flux densities in the strongest pulse and in the average profile indicate that they can be pulsars with giant pulses. One new rotating radio transient (RRAT) J0812+8626 with a dispersion measure $DM = 40.25$~pc/cm$^3$ was detected.

\end{abstract}




\section{Introduction}
In 2006, a new class of pulsars was discovered – rotating radio transient (RRAT) \citeauthor{McLaughlin2006} (\citeyear{McLaughlin2006}). These pulsars emit irregularly appearing (sporadic) dispersed pulses. Regular (pulsar) radiation of rotating transients is often not detected. Over the past 15 years since the discovery of RRAT, approximately one hundred rotating transients have been detected during new surveys and re-processing of data from early surveys. Basically, RRAT are detected in pulsar search surveys (see ATNF pulsar catalog https://www.atnf.csiro.au/people/pulsar/psrcat / and RRATalog http://astro.phys.wvu.edu/rratalog /). Almost all RRATs have been detected on the world's largest radio telescopes. These are radio telescopes Parks (64 meters), Arecibo (300 meters), Green Bank (100 meters), Pushchino (200$ \times$400 meters). When searching for RRAT, the most important factor is the instantaneous sensitivity of the radio telescope, which depends on the effective area and, consequently, on the geometric dimensions of the antenna.

Until now, the nature of rotating transients and their place among ordinary second pulsars have not been determined. On average, RRATs have periods and magnetic fields larger than those of canonical (ordinary) second pulsars and, depending on $P/\dot P$, they often lie close to the line of death \citeauthor{McLaughlin2009}  (\citeyear{McLaughlin2009}), \citeauthor{Keane2011} (\citeyear{Keane2011}). According to the hypothesis \citeauthor{Popov2006} (\citeyear{Popov2006}), the location of the RRAT on the dependence $P/\dot P$ suggests that they may be an intermediate class between pulsars with strong magnetic fields (magnetars, XINS) and ordinary second pulsars. The distribution of heights above the plane of the Galaxy, the distribution of pulse widths of RRAT is the same as that of ordinary pulsars (\citeauthor{Burke-Spolaor2010}, \citeyear{Burke-Spolaor2010}). The average time between the appearing pulses can be from minutes to tens of hours (\citeauthor{McLaughlin2006} \citeyear{McLaughlin2006}, \citeauthor{Logvinenko2020}, \citeyear{Logvinenko2020}). According to \citeauthor{Meyers2019} (\citeyear{Meyers2019}), the appearance of pulses obeys the Poisson distribution, and according to \citeauthor{Lu2019} (\citeyear{Lu2019}) (Fig.6 in the article), some RRAT have pulse clustering. The pulse energy distribution can be lognormal, the sum of two lognormal distributions, lognormal with a power tail, power, power with a break, that is, the distributions are the same as those of ordinary pulsars (\citeauthor{Keane2010} (\citeyear{Keane2010}), \citeauthor{Cui2017} (\citeyear{Cui2017}), \citeauthor{Mickaliger2018} (\citeyear{Mickaliger2018}), \citeauthor{Meyers2019} (\citeyear{Meyers2019}), \citeauthor{Brylyakova2021} (\citeyear{Brylyakova2021}), \citeauthor{Tyulbashev2021s} (\citeyear{Tyulbashev2021s})). According to some estimates, the number of rotating transients may be twice as large as the population of ordinary second pulsars (\citeauthor{Keane2008}, \citeyear{Keane2008}). An attempt to find an evolutionary relationship between pulsars and rotating transients is made in \citeauthor{Keane2008} (\citeyear{Keane2008}), \citeauthor{Keane2010} (\citeyear{Keane2010}).

Unfortunately, the small number of detected RRATs and the difficulties with their investigation due to the sporadic appearance of pulses do not allow us to give an unambiguous answer about the nature of rotating transients. Since the pulse appearance time is unpredictable, and the average time between detected pulses is long, it is necessary to increase the total number of transients and conduct long-term observations lasting tens or hundreds of hours for each point in the sky in order to study the properties of rotating transients and choose the preferred hypothesis about their nature.

In this paper, we consider the search for dispersed pulses in the survey conducted at the radio telescope Large Phased Array (LPA) of the Lebedev Physical Institute (LPI) of the Russian Academy of Sciences.

\section{Observations}

The LPA radio telescope used for the survey is a meridian antenna having four independent beam systems. Two beam systems are used to search for pulsars and transients. One of them (LPA3) has fixed (non-switchable) positions of the beams in height, overlapping declinations from $-9^{\circ}$ to $+55^{\circ}$. LPA3 conducts a daily sky survey used for a number of scientific tasks, including the search for pulsars and RRAT (\citeauthor{Shishov2016} (\citeyear{Shishov2016}), \citeauthor{Tyulbashev2016} (\citeyear{Tyulbashev2016}), \citeauthor{Tyulbashev2018} (\citeyear{Tyulbashev2018})). The other beam system is movable (LPA1), and allows observations at declinations from $-15^{\circ}$ to $+87^{\circ}$. On LPA1, from one to eight beams can be selected for observations, sequentially arranged in a vertical plane according to declinations. The size of one beam is approximately $0.5^{\circ}\times 1^{\circ}$. The BSA1 radio telescope was used in this work. Another beam system is used to monitor the condition of the antenna. For the latter beam system, the possibility of creating an additional full-fledged multibeam radio telescope consisting of beams with fixed coordinates overlapping declinations from $+55^{\circ}$ to $+87^{\circ}$ is being considered. Some details about the capabilities of the LPA antenna after its reconstruction can be found in \citeauthor{Shishov2016} (\citeyear{Shishov2016}). The central frequency of observations is 110.3 MHz, the band is 2.5 MHz.

To conduct the survey on LPA1, a new recorder based on a programmable logic integrated circuit (PLIC) was developed, which is an analogue of the 96-channel recorder already operating on the LPA3 radio telescope. Structurally, the previously developed receivers is a set of modules. In total, there are 12 modules in the recorder, each of which serves 8 beams of the LPA LPI. The modules are installed on the PCI bus of two industrial computers (PCs). The principles of construction, the element base, and the software for conducting observations of this recorder are described in \citeauthor{Logvinenko2020} (\citeyear{Logvinenko2020}).

In the algorithm of the new two-module recorder, the possibility of controlling the transfer of data from the hardware to the computer's random access memory (RAM) via the PCI bus in the mode of the direct access channel (DAC) was introduced. To do this, changes were made to the firmware of the PLIS modules of the recorder and to the program for conducting radio astronomical observations. Thanks to these changes in the recorder, it became possible to use the mode of selective operation of individual modules and, accordingly, the mode of selective data transmission. This reduced the load on data transmission and registration channels and made it possible to use a computer widely used in industry and a conventional OS operating system (Windows 10 PRO). This mode ensured stable collaboration of both the observation program and standard OS system services, as well as the ability of the operator to perform the functions of monitoring the progress of the experiment.

The observation program is integrated into the pulsar observation system at the LPA radio telescope. The operator's participation in the observation process is minimized. New drivers have been developed for the recorder modules to work under Windows 10 64-bit OS. To do this, Microsoft software for driver development was used – Windows driver kit (WDK). When checking the operation of the recorder in hard operating modes, no errors were detected in data registration with simultaneous operation of 2 modules with a time resolution of 3.072 ms and a spectral resolution of 19.53125 kHz (128 spectral channels). If the time resolution is changed, for example, by a factor of 2 to 6.144 ms, the spectral resolution can be increased to 256 spectral channels. The time and frequency resolution parameters are limited by the speed of writing data to the hard disk, the amount of RAM allocated in the system area and the limited computer resources when the OS system services, monitoring and service programs work together.

Declinations of $+56^{\circ}<\delta <+87^{\circ}$, unavailable for observations on LPA3, were selected for the survey. The survey was conducted in eight spatial beams occupying approximately three degrees of declination. The Program Committee for the LPA LPI allocated 9 or 10 days per month for the survey. We fixed a set of declinations for survey on all the selected days, and new declinations were selected for the next month. Thus, a one-time survey of declinations covering $31^{\circ}$ was conducted over 10 months.

Observations were started in April 2019. By the end of December 2020, each point in the sky with declinations between $+56^{\circ}$ and $+87^{\circ}$ had at least 9 observation sessions, which corresponds to continuous observations of at least one time hour. The viewing area is about 4000 sq.deg. Basically, these are observations at high galactic latitudes and in the anti-center of the Galaxy.

The recording was carried out in a 2.5 MHz band, divided into 128 frequency channels with a channel width of 19.53 kHz. The selected readout time was 3.072 ms. Observations were carried out around the clock. One file was recorded every hour. 24 files are recorded per day. The calibration signal, common to all radio telescopes implemented on the LPA LPI antenna, is supplied 6 times a day. It looks like an OFF-ON-OFF (calibration step). At a given time, the antenna is turned off for five seconds (OFF mode), then when the antenna is turned off, the calibration signal of a known temperature is turned on for five seconds (ON mode), then the calibration signal is turned off, and another five-second recording is performed. After recording the calibration step, the antenna turns on. The fixed height of the calibration step in units of an analog-to-digital converter (ADC) has a strong dependence on the ambient temperature and varies both in different seasons and during the day (for more information, see \citeauthor{Tyulbashev2019} (\citeyear{Tyulbashev2019})). In this paper, the calibration step was used to equalize the gain in the frequency channels.

\section{Results}

The search for dispersed pulses was carried out using a specially developed program written in the C$\#$ language in the Microsoft Visual Studio environment. Data processing and the search for pulse events with signs of dispersion lag are carried out in several stages. At the first stage, for the set of variance measures specified in the processing parameters, all events exceeding the specified value of the signal-to-noise ratio (S/N) are recorded and recorded. Next, a sequential analysis is performed in order to reject interference and repeat detections of the same event. Events caused by pulse interference are eliminated. Temporary sections corresponding to the calibration step record are rejected. During processing, over a long time interval (several days, weeks), time intervals corresponding to signals from already identified space objects, including signals in the side lobes and in neighboring beams, are determined. Such areas, by the decision of the operator, can be entered into a table located in external files, and do not participate in processing. From the remaining set of events, groups are selected that may be the result of the presence of a single signal with a dispersion delay in the recording. For each of these groups, only one event (a graphic file) with the highest S/N value is allocated and left. The graphic file contains the pulse profile, the dynamic spectrum of the pulse, as well as other technical parameters of the event: the name of the file in which the detected pulse is recorded, the pulse width, the DM estimate, the coordinates of the pulse in right ascension, the coordinates of the BSA beam in declination, and others. These make it possible to reduce the number of events requiring further manual processing. Events that have passed the cull are recorded in folders corresponding to the beam and sidereal time in which they are recorded.

\begin{figure}
\begin{center}
	\includegraphics[width=0.9\columnwidth]{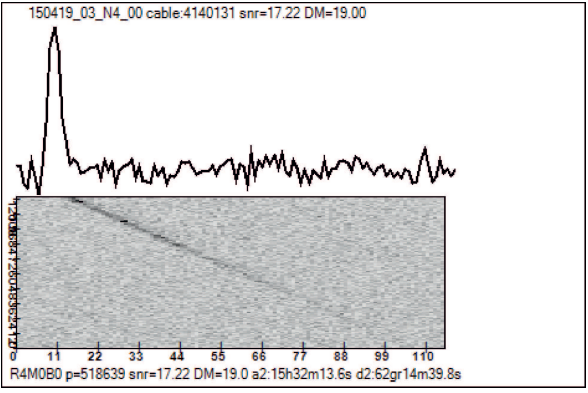}
    \caption{An example of a service drawing created by the search program for each detected pulse. The upper part of the figure shows the profile of one pulse of the pulsar J1509+5531 detected in the side lobe of the LPA LPI in the direction $\alpha_{2000} = 15^h32^m; \, \delta_{2000} = 62^{\circ}15^\prime$. The dynamic spectrum of the pulse is shown in the lower part of the figure. The horizontal axis shows the duration of the recording in points. 110 points correspond to approximately 338 ms of raw data recording. The vertical axis of the dynamic spectrum reflects the frequencies. The top of the dynamic spectrum corresponds to the frequency of 111.49 MHz, and the bottom to the frequency of 109.01 MHz}
    \label{fig:fig1_search}
\end{center}
\end{figure}

During the blind search, 75 directions in the sky were detected, from which dispersed signals are observed. The analysis showed that some of the pulses belong to known pulsars observed in the main beam, some of the pulses are associated with known pulsars observed in the side lobes of the antenna, some of the pulses remained unidentified.

The pulses of known pulsars detected in the side lobes of the LPA LPI present a serious problem in identification. If several pulses are detected on a given day, then in addition to $DM$, it is possible to roughly estimate the upper estimate of the period and, accordingly, obtain possible multiples of periods. The obtained estimates of the period and $DM$ allow us to check the strong pulsars included in the review as candidates for identification. On the other hand, it may turn out that a strong pulsar is located outside the viewing area, and there is no estimate of the period of the detected transient. We cannot guarantee the absence of fake detection of new transients due to the complex distribution of the side lobes of the LPA LPI, however, we are making every possible effort to exclude such cases. In the side lobes of the antenna, we detected 13 pulsars located outside the viewing area. For example, the pulsar J1509+5531, located outside the investigated site, was detected 333 times in the side lobes of the LPA LPI. Pulsar pulses were detected from 1 to 101 times in different directions (see Figure 1 for an example of the PSR J1509+5531 pulse). In some of the known pulsars located in the viewing area, pulses in the side lobes are also detected.

\begin{table*}
	\centering
	\caption{Characteristics of discovered pulsars.}
	\label{tab:example_table}
	\begin{tabular}{ccccccccccc} 
		\hline
		
$name_{LPA}$ & $name_{ATNF}$ & N & $S_{p1}$ (Jy) & $W_{e1}$ (ms) & $S_{p2}$ (Jy) & $W_{e2}$ (ms) & $S_{int}$ (mJy) & $S_{135}$ (mJy) & $S_{102}$ (mJy) & $S_{p1} / S_{p2}$\\
		\hline
J0140+6008* & J0141+6009 & 562 & 241.5 & 12.3 & 14.3 & 33.8 & 394 & 102.6 & -&16.9\\
J0158+6223&J0157+6212&2&17.7&12.3&0.65&64.5&17.8&4.8&52&27.2\\
J0653+8054*&J0653+8051&15&15.5&6.1&0.7&27.6&16&13.1&16&22.1\\
J0750+5724&J0750+57(St)&1&>14&6.1&>0.27&24.6&>6&-&-&-\\
J0814+7436*&J0814+7429&62 &781.4&6.1&56.3&61.4&2674&358.8*&1080&13.9\\
J1058+6504&J1059+6459(St)&3&4.2&12.3&0.62&33.8&5.8&-&-&6.8\\
J1708+5858&J1706+59(St)&24&>9.0&6.1&>0.55&58.4&>22&15.7&-&-\\
J1843+5640*&J1840+5640&757&235.4&6.1&21.4&21.5&275&55.0&50&11\\
J1911+5654&J1910+5655(S)&2&28.5&3.1&0.26&61.4&46.7&-&-&109\\
J2225+6527*&J2225+6535&94&55.5&3.1&6.0&27.6&242&126.3&-&9.3\\
J2336+6145*&J2337+6151&346&77.6&9.2&2.4&27.6&132.4&28.7&75&32.7\\
J2354+6158&J2354+6155&2&50.6&12.3&1.6&15.4&26&10.5&30&31.6\\		
		\hline
	\end{tabular}
	\label{tab:tab1}
\end{table*}

\begin{figure}
\begin{center}
	\includegraphics[width=1.0\columnwidth]{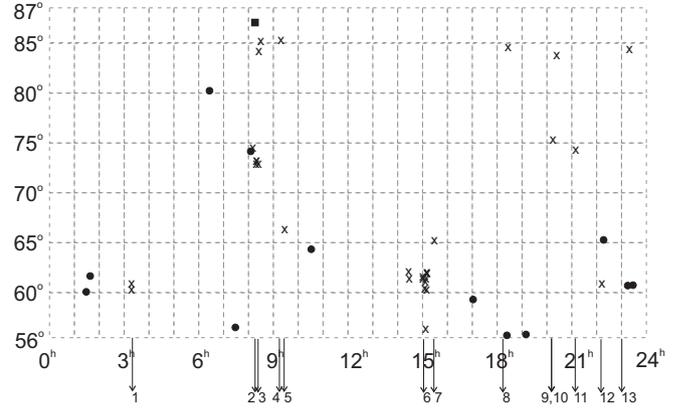}
    \caption{The figure schematically shows the viewing area, where the coordinates of the detected dispersed pulses are marked. Horizontally, the coordinates for right ascension are postponed, and vertically, the coordinates for declination. The crosses mark the observations of known pulsars in the side lobes. The numbers 1-13 indicate pulsars whose coordinates are located outside the studied area: 1 – J0332+5434; 2 – J0826+2637; 3 – J0837+0610; 4 – J0922+0638; 5 – J0946+0951; 6 – J1509+5531; 7 – J1543+0929; 8 – J1823+0550; 9 – J2022+2854; 10 – J2022+5154; 11 – J2113+2754; 12 – J2219+4754; 13 – J2305+3100. The shaded circles show the locations of 12 pulsars detected in the main beam. The shaded square shows the detected new rotating radio component.}
    \label{fig:fig2_search}
\end{center}
\end{figure}

Fig.~\ref{fig:fig2_search} shows the coordinates of the directions of the dispersed pulses detected by us. Crosses indicate 13 pulsars found in the side lobes and located outside the investigated site. It can be seen that the side lobes have a complex distribution, but all the detected pulses have a close right ascension with respect to the pulsars that generated the observed pulses. Detections in the side lobes for known pulsars that have fallen into the observed area are not shown, so as not to weigh down the drawing.

If a pulsar is detected in the main beam, then it is easy to identify it. The obtained estimates of $DM$ and coordinates for right ascension and declination allow us to select candidates for identification in ATNF. The subsequent verification of candidates by averaging raw data with a period and a $DM$ taken from ATNF makes the identification unambiguous. If no regular pulsar radiation was detected, the object was placed in the list of RRAT candidates.

Analysis of the detected pulses showed that we observed 12 known pulsars in the main beam and one new RRAT. The remaining 62 sources of pulsed radiation are the radiation of known pulsars in the side lobes of the LPA LPI. For the identified pulsars and the detected RRAT, estimates of the flux density were made using calibration sources. Calibration sources were selected in such a way that their declination coordinates were close to the coordinates of the detected object, and the right ascension differed by no more than two time hours. Candidates for calibration sources were selected from the Pushchino catalog of discrete sources \citeauthor{Dagkesamanskii2000} (\citeyear{Dagkesamanskii2000}) (http://astro.prao.ru/db/), which was made at a frequency of 102.5 MHz. In the work \citeauthor{Tyulbashev2019} (\citeyear{Tyulbashev2019}) it is shown that when recalculating the flux density of the calibration source to a frequency of 111 MHz, the corrections will be insignificant. Assuming that the spectral indices $\alpha$ ($S\sim\nu^{-\alpha}$) for all calibration sources are equal to one, the flux densities at 111 MHz will be equal to 0.94 of the flux densities at 102.5 MHz. B0245+603 (44.3 Jy), B0733+806 (32.3 Jy), B0735+744 (12.6 Jy), B0742+576 (14.2 Jy), B1107+651 (12.0 Jy), B1656+572 (13.6 Jy), B1752+586 (10.0 Jy), B1858+568 (17.4 Jy, B2159+652 (28.1 Jy), B2356+620 (27.2 Jy) were selected as calibration sources).

Table~\ref{tab:tab1} shows the characteristics of the detected pulses for identified pulsars. In the first and second columns of the table are the name of the source in the J2000 annotation according to the detection on the LPA LPI and according to the identification in the ATNF catalog. Since the radiation pattern of the LPA LPI is of the order of a degree, the accuracy of our coordinates is low, and therefore the names may not match. We give both names so that the accuracy of the coordinates can be compared. In the third column, the number of pulses have detected in the main beam. Columns 4-7 contain the peak flux density of the strongest detected pulse ($S_{p1}$), the half-width of this pulse, the peak flux density in the average profile ($S_{p2}$), the half-width of the average profile. The average profile was built during the same session when the strongest pulse was observed. The strongest impulse itself was not excluded when constructing the average profile. Estimates show that the exclusion of a strong impulse will lead to a decrease in $S_{p2}$ by $15-20\%$. Columns 8-10 contain estimates of the integral flux density at 111 MHz ($S_{int}$), at 135 MHz (\citeauthor{Sanidas2019}, \citeyear{Sanidas2019}) and at 102.5 MHz (\citeauthor{Malofeev2000}, \citeyear{Malofeev2000}). When obtaining estimates of the flux density at 111 MHz, it was taken into account that the position of the LPA beams in the sky is fixed, and the pulsar can be located above or below the beam. Therefore, corrections of the flux density were made, taking into account the features of the LPA LPI, as an antenna array. For pulsars J0750+57 and J1706+59, discovered in 2014 (\citeauthor{Stovall2014}, \citeyear{Stovall2014}), the coordinates of pulsars with low accuracy are given in the ATNF catalog. Therefore, no correction was made for the mismatch of the beam coordinate and the pulsar coordinate. For these pulsars, our estimate of the flux density may be underestimated by several times. Column 11 contains the quotient of columns 4 and 8, that is, the value of how many times the peak flux density of the strongest pulse differs from the peak flux density in the average profile for the same day.

The asterisk in the first column marks pulsars observed both in the main beam and in the side lobes. The labels "St" (\citeauthor{Stovall2014}, \citeyear{Stovall2014}) and "S" (\citeauthor{Sanidas2019}, \citeyear{Sanidas2019}) in the second column indicate the work with the first detection of these pulsars. These pulsar search works were going on at about the same time as our review on the search for new pulsars and transients. We confirm the detection of these pulsars at a frequency of 111 MHz. The asterisk in the ninth column shows the expected flux density of the pulsar J0814+7429 at a frequency of 135 MHz. We also specify the value of the dispersion measure for pulsars J0750+57 ($DM = 26.75\pm 0.25$pc/cm$^3$) and J1706+59 ($DM = 30.5 \pm 0.25$pc/cm$^3$), determined by us from the strongest observed pulses.

\begin{figure*}
\begin{center}
	\includegraphics[width=1.0\textwidth]{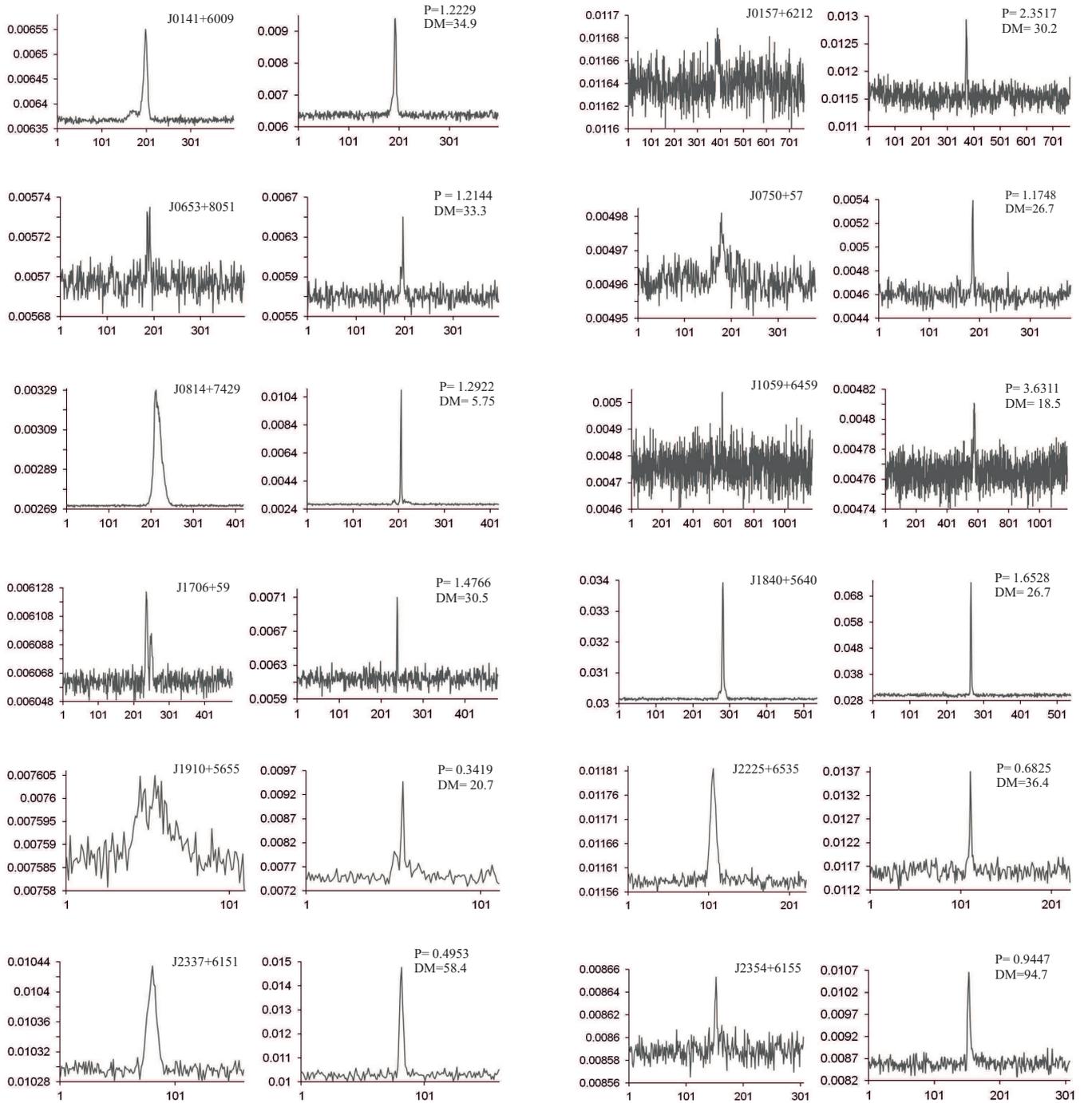}
    \caption{The paired figures on the left show the average profiles of known pulsars detected in the survey by individual pulses, and on the right are the profiles of the strongest detected pulses of these pulsars. On the horizontal axis, digitization in points (3.072 ms per point). The full period is given for all pulsars. The vertical axis shows the flux density in conventional units. Conventional units for paired drawings are given in one scale, which allows you to see how many times the peak flux densities of the strongest pulses are higher than the peak flux densities in the average profiles. The name of the pulsar is shown on the left side of the paired figure, its period and the measure of dispersion are shown on the right side of the figure.}
    \label{fig:fig3_search}
\end{center}
\end{figure*}

Fig.~\ref{fig:fig3_search} shows the average profiles and strongest pulses of known pulsars found during the survey. It can be seen that the half-widths of the strongest pulses are significantly smaller than the half-widths of the average pulsar profiles. Using columns 5 and 7 in Table 1, it is possible to estimate how many times the strongest impulse is narrower than the average profile. The greatest difference in the half-width of the pulse and the average profile of the pulsar is J1910 +5655, where the half-widths differ by 19.8 times. The smallest difference in the half-widths of the pulsar is J2354+6155, where the half-widths of the average profile and the strongest pulse differ by 1.3 times. The median value of the half-width difference falls on values 4-5.

According to column 11 in Table~\ref{tab:tab1}, the peak flux density of the strongest pulses is usually ten or more times greater than the peak flux density in the average pulsar profile. According to the study of pulsars with giant pulses conducted at the LPA LPI (\citeauthor{Kazantsev2018}, \citeyear{Kazantsev2018}), there are a number of signs that distinguish a pulsar with giant pulses from a conventional pulsar. Two signs that can be checked according to our data are the small width of the giant pulse in comparison with the average profile and the difference between the peak flux density of the giant pulse and the integral flux density of 30 (strong criterion) or 10 (weak criterion) or more times. Of the 12 pulsars in Table~\ref{tab:tab1}, such an estimate could not be made for two pulsars, and for two more, the peak flux densities differ by less than 10 times. Eight pulsars satisfy the weak criterion, and four of these eight satisfy the strong criterion. Therefore, four pulsars are good candidates for pulsars with giant pulses.

Separate studies are required in order to check whether the sources found by dispersed pulses are pulsars with giant pulses. In particular, one of the objects (B0809+74/J0814+7429) was specially studied in \citeauthor{Kazantsev2018} (\citeyear{Kazantsev2018}) as a candidate for pulsars with giant pulses. For the pulsar , almost $2.8\times 10^5$ pulses, of which 49 met the criterion of "gigantism", namely, their peak flux density of 1500-2000 Jy was 30 or more times higher than the peak flux density in the average profile. The energy distribution of the pulses turned out to be lognormal with a power tail. According to the authors, the energy distribution of pulses constructed by them and the sub-pulse drift observed for this source indicate that the pulsar J0809+7429 has abnormally strong, but not gigantic pulses.

In addition to the known pulsars, a source was found that could not be identified. In J0812+8626, indicated in Fig.2 by a shaded rectangle, two pulses were detected at $DM = 40.25 \pm 0.25$ pc/cm$^3$. The distance between the pulses turned out to be 47.58s. The search for periodic radiation in the direction of the source is performed in the summed power spectra and in the summed periodograms. Since we had at least 9-10 days of observations in the direction of each source, the incoherent addition of the power spectra (\citeauthor{Tyulbashev2017} (\citeyear{Tyulbashev2017}), \citeauthor{Tyulbashev2020} (\citeyear{Tyulbashev2020})) and periodograms  (\citeauthor{Tyulbashev2021p}, \citeyear{Tyulbashev2021p}) over all the days of observations allowed us to increase the sensitivity when searching for periodic radiation by about 2-3 times. The peak flux density of the found pulses is $S_p= $10 and 4.5 Jy, the half-width of the profile is $W_e = 10$~ms. The upper estimate of the flux density is $S_{int} < 2$~mJy (if $0.5c < P < 10c$). Since the appearance of RRAT pulses is sporadic, the true coordinates of the source can come to any place in the radiation pattern. We give the formal coordinates of J0812+8626: $\alpha_{2000}= 08^h12^m30^s \pm 2.5^m; \, \delta_{2000}= 86^{\circ}26^\prime \pm 15^\prime$. The exact coordinate of the transient is not known, so it is not possible to make corrections to the flux density that take into account the position of the LPA LPI beams relative to the source. The given values of peak flux densities are lower estimates. The actual flux density of these pulses can be 1.5-2 times higher.

\begin{figure}
\begin{center}
	\includegraphics[width=1.0\columnwidth]{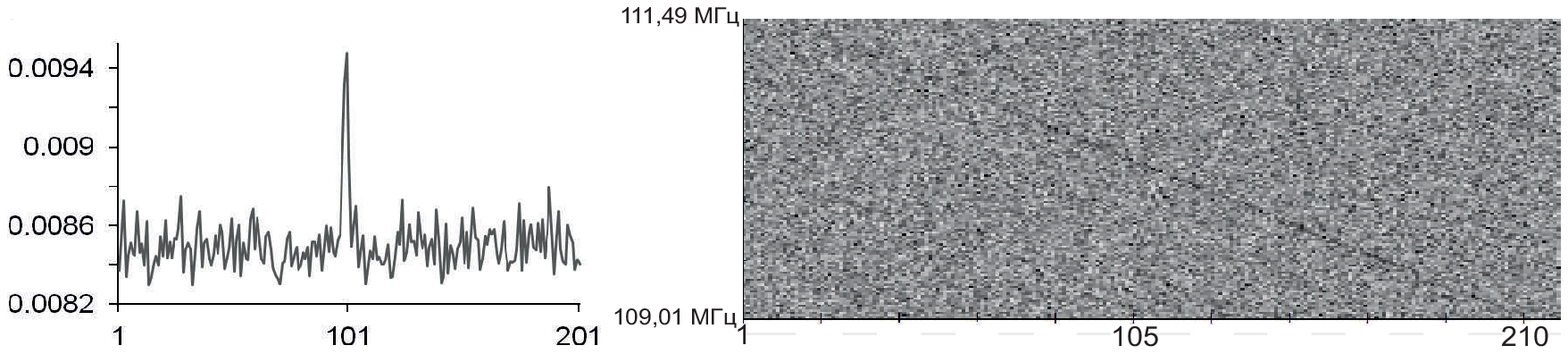}
    \caption{The figure shows the dynamic spectrum and profile of the strongest pulse RRAT J0812+8626. The vertical axis of the dynamic spectrum shows the frequencies of observations. For the profile of a single pulse along the vertical axis, the flux density is given in conventional units. The horizontal axes show the observation time in points (one point is 3.072 msec).}
    \label{fig:fig4_search}
\end{center}
\end{figure} 

\section{Discussion of the results}

When searching for dispersed pulses in a area covering approximately 4000 sq. deg. 13 sources of pulsed radiation were detected. 12 of them turned out to be known pulsars, which detected from 1 to 757 pulses. Four pulsars (J0750+57; J1059+6459; J1706+59; J1910+5655) out of twelve were found in recent searches for pulsars at a frequency of 1400 MHz (Green Bank, \citeauthor{Stovall2014}, \citeyear{Stovall2014}) and at a frequency of 135 MHz (LOFAR, \citeauthor{Sanidas2019}, \citeyear{Sanidas2019}), and we confirm them detection at a frequency of 111 MHz. For two (J0750+57; J1706+59) we have refined the estimate of $DM$. A comparison of the half-widths of the strongest pulses of all 12 pulsars and the half-widths of the average profiles of these pulsars shows that the average profiles are from 1.5 to 20 times wider than the strongest pulses. Comparison of peak flux densities of the strongest pulses and average profiles shows that four pulsars (J0157+6212; J1910+5655; J2337+6151; J2354+6155) are good candidates for pulsars with giant pulses. The best candidate is the pulsar J1910+5655. The peak density of the flux of its strongest pulse exceeds the peak density of the flux of the average profile by 109 times. The half-widths of the strongest pulse and the average profile differ by 20 times.

One of the hypotheses about the nature of RRAT was expressed by \citeauthor{Weltevrede2006} (\citeyear{Weltevrede2006}). According to the proposed hypothesis, rotating transients are ordinary pulsars with an unusually long tail in the energy distribution of pulses. It is assumed that the integral flux density of these pulsars is below the detection threshold for this radio telescope, and individual pulses from the tail of the distribution are strong enough to detect them. The maximum sensitivity for LPA LPI when searching for periodic radiation is 5-10 mJy (\citeauthor{Tyulbashev2016}, \citeyear{Tyulbashev2016}), the maximum sensitivity of LPA LPI when searching for pulsed radiation is approximately 2 Jy (\citeauthor{Tyulbashev2018}, \citeyear{Tyulbashev2018}). Let's select the numbers, when multiplied by which the integral density of the pulsar flux will decrease to 5 mJy, and multiply by the selected number the peak flux density of the strongest pulse of this pulsar. We get: J0141+6009 (3.1 Jy); J0157+6212 (5.0 Jy); J0653+8051 (4.8 Jy); J0814+7429 (1.5 Jy); J1059+6459 (3.6 Jy); J1840+5640 (4.3 Jy); J1910+5655 (3.1 Jy); J2225+6535 (1.1 Jy); J2337+6151 (2.9 Jy); J2354+6155 (9.7 Jy). Thus, out of 10 pulsars with estimated peak flux density, when the pulsar is removed to such a distance that its integral flux density drops to 5 mJy, individual pulses will not be visible only in two pulsars (J0814+7429; J2225+6535).

The hypothesis \citeauthor{Weltevrede2006} (\citeyear{Weltevrede2006}) may be valid for part of the sample of ordinary second pulsars. At the same time, as shown in \citeauthor{McLaughlin2009}  (\citeyear{McLaughlin2009}), \citeauthor{Keane2011} (\citeyear{Keane2011}), typical periods of RRAT are many times larger than typical periods of second pulsars. Therefore, some of the observed rotating transients belong to some other sample. It is possible that part of the RRAT sample is associated with giant pulses. In the works \citeauthor{Brylyakova2021} (\citeyear{Brylyakova2021}) and \citeauthor{Tyulbashev2021s} (\citeyear{Tyulbashev2021s}) for several RRAT, it was possible to show that the observed pulses can be giant pulsar pulses.

Two pulses of the new RRAT J0812+8626 were detected. The detected number of pulses allows us to speak only about the detection of RRAT, but does not make it possible to conduct any analysis.

\bsp	
\label{lastpage}
\end{document}